\documentclass[prl,twocolumn,superscriptaddress]{revtex4}

\usepackage{bm}
\usepackage{graphics}
\usepackage{amsmath,epsfig}
\usepackage{amssymb}
\usepackage{comment}
\begin{document}

\title{Demonstration of Shor's quantum factoring algorithm using photonic qubits}

\author{Chao-Yang Lu $^\clubsuit$}
\affiliation{Hefei National Laboratory for Physical Sciences at
Microscale and Department of Modern Physics, University of Science
and Technology of China, Hefei, Anhui 230026, P. R. China}

\author{Daniel E. Browne $^\spadesuit$}
\affiliation{Department of Materials and Department of Physics,
Oxford University, Parks Road, Oxford, OX1 3PU, United Kingdom}

\author{Tao Yang $\ddag$}
\affiliation{Hefei National Laboratory for Physical Sciences at
Microscale and Department of Modern Physics, University of Science
and Technology of China, Hefei, Anhui 230026, P. R. China}

\author{Jian-Wei Pan $^\P$}
\affiliation{Hefei National Laboratory for Physical Sciences at
Microscale and Department of Modern Physics, University of Science
and Technology of China, Hefei, Anhui 230026, P. R. China}
\affiliation{Physikalisches Institut, Universit\"{a}t Heidelberg,
Philosophenweg 12, 69120 Heidelberg, Germany}

\date{Submitted to Phys. Rev. Lett. on 17 April}

\begin{abstract}
We report an experimental demonstration of a complied version of
Shor's algorithm using four photonic qubits. We choose the simplest
instance of this algorithm, that is, factorization of $N=15$ in the
case that the period $r=2$ and exploit a simplified linear optical
network to coherently implement the quantum circuits of the modular
exponential execution and semi-classical quantum Fourier
transformation. During this computation, genuine multiparticle
entanglement is observed which well supports its quantum nature.
This experiment represents a step toward full realization of Shor's
algorithm and scalable linear optics quantum computation.
\end{abstract}

\pacs{Valid PACS appear here}

\maketitle

Shor's algorithm \cite{shor1,shor2} for factoring large numbers is
arguably the most prominent quantum algorithm to date. It provides a
way of factorizing large integers in polynomial time using a quantum
computer, a task for which no efficient classical method is known.
Such a capacity would be able to break widely used cryptographic
codes, such as the RSA public key system \cite{RMP96,rsa}.
Experimental realization of Shor's algorithm has been a central goal
in quantum information science. Owing to its high experimental
demands, however, this algorithm has so far only been demonstrated
in a nuclear magnetic resonance (NMR) experiment \cite{LMKV}. Since
the NMR experiments cannot prepare pure quantum states and exhibits
no entanglement during computation, concerns have been arisen on its
quantum nature \cite{nature}. In particular, it has been proved that
the presence of entanglement is a necessary condition for quantum
computational speed-up over classical computation \cite{nature}.

The approach of using photons to implement quantum algorithms is
appealing due to the long decoherence times and precise single-qubit
operations \cite{KLM,nielsen,rmp}. Along this line, the
Deutsch-Josza algorithm and Grover algorithm have been realized (see
e.g. \cite{steinberg,tame,kwiatgrover,walther,robert}). In this
Letter, we use the photonic qubits to demonstrate the easiest
meaningful instance of Shor's algorithm, that is, factorization of
$N=15$ in the case that the period $r=2$. A simplified linear optics
network is designed to implement the quantum circuit. We have
observed genuine multiparticle entanglement and multipath
interference during computation, which thus for the first time prove
the quantum nature of the implementation of Shor's algorithm
\cite{nature}.

Suppose we wish to find a non-trivial prime factor of an $l$-digit
integer $N$. Even using the best known classical algorithm, prime
factorization takes exponentially many operations, which quickly
becomes intractable as $l$ increases. Shor's algorithm, in contrast,
offers a new powerful way to solve this problem in only polynomial
time \cite{shor1, shor2}. The strategy for the quantum factoring of
a composite number $N=p\,q$, with both $p$ and $q$ being odd primes,
is as follows. First we pick a random number $a$ ($0<a<N$) with no
factor in common with $N$. Then we quantum compute the period $r$ of
the modular exponential function (MEF): $f(x)=a^x \textrm{mod} N$,
which is the smallest positive satisfying $a^r\textrm{mod} N=1$.
From this period $r$, at least one nontrivial factor of $N$ is given
by the greatest common denominator ($\mathrm{g.c.d.}$) of
$a^{r/2}\pm1$ and $N$ with probability greater than 1/2.

Shor's algorithm provided an efficient quantum circuit (see Fig.~1a)
to find the period $r$. Generally two registers of qubits are used
\cite{shor1,preskill}; the first register  with
$n=2\lceil\log_2N\rceil$ qubits and the second one with
$m=\lceil\log_2N\rceil$ qubits. Applying Hadamard (H)
transformations on the first register which was initialized in the
state $|0\rangle^{\otimes n}$, it becomes
$2^{-n/2}\sum_{x=0}^{2^n-1}|x\rangle$, an equally weighted coherent
superposition of all $n$-qubit computational basis. Then the MEF
unitarily implements $a^x\mathrm{mod}N$ on the second register when
the first register is in state $|x\rangle$, giving
\begin{equation}\label{2}
 \frac{1}{\sqrt{2^n}}\sum_{x=0}^{2^n-1}|x\rangle |a^x
\textrm{mod} N\rangle.
\end{equation}
The highly entangled state Eq. (\ref{2}) exhibits what Deutsch
called ``massive quantum parallelism'' \cite{preskill, Deutsch}, as
although the execution has run for only once, it entangles all the
$2^n$ input value $x$ with the corresponding value of $f(x)$ in
parallel. Next, the quantum Fourier transformation (QFT) is applied
on the first register, yielding
\begin{equation}\label{3}
 \frac{1}{2^n}\sum_{y=0}^{2^n-1}\sum_{x=0}^{2^n-1}e^{2\pi i\,xy/2^n}|y\rangle\:|a^x
\textrm{mod} N\rangle,
\end{equation}
where interference leads to peaks in the probability amplitudes
for terms $|y\rangle$ with $y=c\,2^n/r$ (for integer $c$). Thus
the period $r$ can be deduced with high probability \cite{shor1}.
The QFT on $2^n$ elements can be efficiently performed on a
quantum computer with $O(n^2)$ gates.
\begin{figure*}[htb]
\centering
  \includegraphics[width=0.98\textwidth]{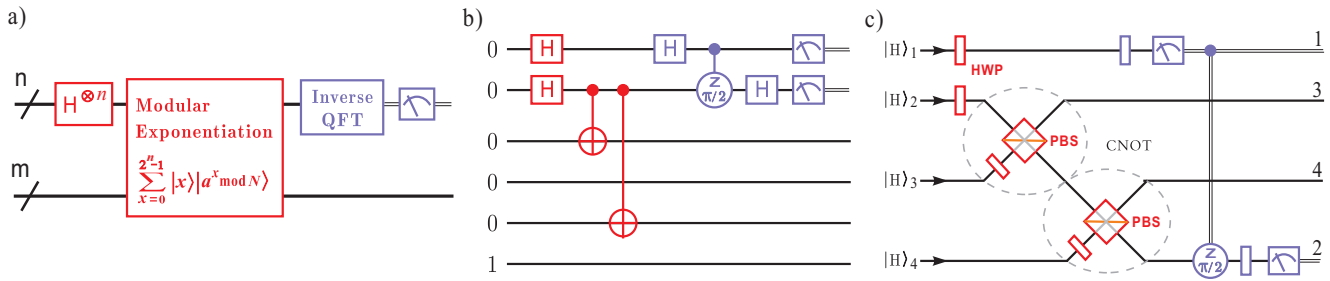}\\
  \caption{Quantum circuit for the order-finding routine of Shor's algorithm.
  (a). Outline of the quantum circuit.
  (b). Quantum circuit for $N=15$ and $a=11$.
  The MEF is implemented by
  two CNOT gates and the QFT is implemented
  by Hadamard rotations and two-qubit conditional
  phase gates. The gate-labeling scheme denotes the axis
  about which the conditional rotation takes place
  and the angle of rotation. (c). The simplified linear optics network using
  HWPs and PBSs to implement the MEF circuit
  and the semiclassical version of the QFT circuit.
  The double lines denote classical information.
     }\label{}
\end{figure*}

Implementations of this algorithm, even for factorization of a small
number, place a lot of challenging experimental demands, e.g.,
coherent manipulations of multiple qubits and creations of
highly-entangled multiqubit registers. Here we aim to demonstrate
the simplest instance of Shor's algorithm, i.e., the factorization
of $15$. Quantum networks for evaluating the MEF have been designed
which involve $O(n^3)$ operations \cite{preskill, vedral}. Since
$a^x=a^{2^{n-1}x_{n-1}}\cdots a^{2x_1}a^{x_0}$, the execution of MEF
can be decomposed into a sequence of controlled multiplications. A
general purpose algorithm to factorize $15$ would require at least
$n=8,$ $m=4,$ thus total $12$ qubits \cite{preskill}. Several
observations allow us to reduce the resources substantially for the
purpose of a proof-of-principle demonstration. First we choose to
implement the algorithm with $a=11$, this was identified in
\cite{LMKV} as the ``easy'' case. Since $a^2\mathrm{mod}15=1$, MEF
can be simplified to multiplications controlled only by $x_0$, which
can be implemented by two controlled-NOT (CNOT) gates
\cite{Lie-thesis}. A QFT then follows to read out the period $r$.
Such a circuit is shown in Fig.~1b. We note there are two qubits in
the second register which evolve trivially during computation and
can thus be left out.

To demonstrate the circuit of Fig.~1b we use single photons as
qubits, where $|0\rangle$ and $|1\rangle$ are encoded with the
photon's horizontal ($H$) and vertical ($V$) polarization
respectively. The difficulty in implementing this circuit lies in
the CNOT gates and conditional $\pi/2$-phase shift gate. Although
such entangling gates are possible for photons in principle using
measurement-induced nonlinearity \cite{KLM}, currently they are
still experimentally expensive \cite{1cnot,rmp}. Here we note that
since the target qubits of the CNOT gates are always fixed at
$|H\rangle$, so the gate could be realized in an easier and more
efficient fashion. Such a CNOT gate use only a polarizing beam
splitter (PBS) and a half-wave plate (HWP), through which an
arbitrary control qubit $(\alpha|H\rangle+\beta|V\rangle)$ and the
target qubit $|H\rangle$ evolve into
$\alpha|H\rangle|H\rangle+\beta|V\rangle|V\rangle$ upon
post-selection \cite{dik}, that is, conditioned on that there is one
and only one photon out of each output (see Fig.~1c). Furthermore,
the QFT circuit can also be implemented with a more efficient
method. It was observed by Griffiths and Niu \cite{griffiths} that
when immediately followed by measurements, the fully coherent QFT
can be replaced by a semiclassical version that employs only
single-qubit rotations conditioned on measurement outcomes. This
eliminates the need for entangling gates and reduces the numbers of
gates quadratically. Thus we finally arrive at the simplified linear
optics MEF and QFT network in Fig.~1c. We note despite of these
simplifications, our circuit suffices to demonstrate the underlying
principles of this algorithm.

Now we proceed with the experimental demonstration. Our experimental
set-up is illustrated in Fig.~2, where a pulsed ultraviolet laser
passes through two $\beta$-barium borate (BBO) crystals to create
two pairs of entangled photon \cite{Kwiat}. We use polarizers to
disentangle the photons and prepare them in the states $|H\rangle_i$
with $i$ denoting the spatial modes (see Fig.~1c). The photons pass
through the HWPs and are superposed on the PBSs (see Fig.~2) to
implement the necessary single- and two-qubit gates. To ensure good
spatial and temporal overlap, the photons are spectrally filtered
($\mathrm{\Delta\lambda}_{\mathrm{\,FWHW}}=3.2\,\mathrm{nm}$) and
coupled by single-mode fibers \cite{zukowski}.

\begin{figure}[hbt]
\centering
  \includegraphics[width=0.48\textwidth]{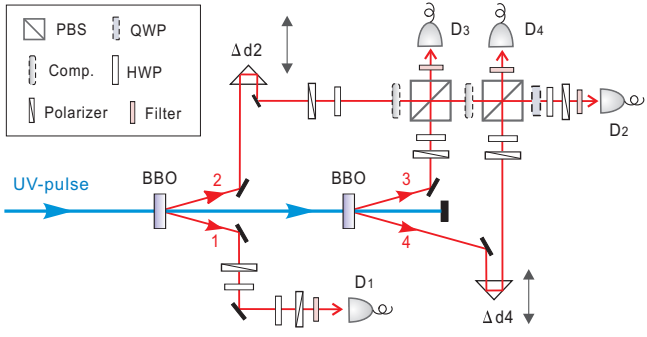}\\
  \caption{Experimental set-up.
  Femtosecond laser pulses (394nm, 120fs, 76MHz) pass through two BBO crystals to produce
  two pairs of
  entangled photons with an average count of $7.8\times10^4s^{-1}$.
  Fine adjustments of the delays between path 2, 3, 4 are
  made by translation stages $\mathrm{\Delta d_2}$ and $\mathrm{\Delta d_4}$.
  We incorporate in front of each PBS a compensator (Comp.) to counter
  the additional phase shifts of the PBS. The final measurement results are then read out using polarizers and single-photon detectors.
     }\label{}
\end{figure}

How could one experimentally verify a valid demonstration of Shor's
algorithm? First let us see the theoretical predictions. After
$a=11$ is chosen, the first step of this algorithm, the MEF should
evolve as $(1/2)\sum_{x=0}^{3}|x\rangle |11^x\textrm{mod} 15\rangle
  =(1/2)(|0\rangle|1\rangle+|1\rangle|11\rangle+|2\rangle|1\rangle+|3\rangle|11\rangle).$
As we rewrite it in binary representation
$(|000001\rangle+|011011\rangle+|100001\rangle+|111011\rangle)/2$,
it shows that a nontrivial Greenberger-Horne-Zeilinger (GHZ)
\cite{ghz} entangled state
$|\psi\rangle=(1/\sqrt{2})(|0\rangle_2|0\rangle_3|0\rangle_4+|1\rangle_2|1\rangle_3|1\rangle_4)$
is created between the two registers. For Shor's algorithm as well
as some others, multiqubit entanglement is a necessary condition if
the quantum algorithm is to offer an exponential speed-up over
classical computation \cite{nature}. In our experiment, as the
photons pass through the MEF circuit, we first observe the
Hong-Ou-Mandel type interference \cite{hom} of three photons in arms
$2$-$3$-$4$ (see Fig.~3b). Then, after fixing the delays at the zero
positions, we expect to determine the fidelity of the three-photon
GHZ state and detect its genuine entanglement. The fidelity is
judged by the overlap of the experimentally produced state with the
ideal one: $F_{\psi}=\langle\psi|\rho_{\mathrm{exp}}|\psi\rangle$.
Here $\rho$ can be decomposed as
$\rho=|\psi\rangle\langle\psi|=(1/2)[(|HHH\rangle\langle
HHH|+|VVV\rangle\langle VVV|)+(1/4)(XXX-XYY-YXY-YYX)]$
\cite{witness-exp}, where $X$ and $Y$ denote Pauli matrices
$\sigma_x$ and $\sigma_y$ which correspond to measurements in basis
$|\mathrm{+}/\mathrm{-}\rangle=(1/\sqrt{2})(|H\rangle\pm |V\rangle)$
and $|R/L\rangle=(1/\sqrt{2})(|H\rangle\pm i|V\rangle)$
respectively. Figure 3c-3d shows the measurement results, which
yield $F_{\psi}=0.74\pm0.02$. It is proved that a fidelity above
$50\%$ is sufficient to show genuine entanglement of the GHZ states
\cite{witness-exp}. Thus the presence of genuine entanglement
created between the two registers of our quantum computer is
confirmed by 12 standard deviations.

\begin{figure}[htb]
\centering
  \includegraphics[width=0.49\textwidth]{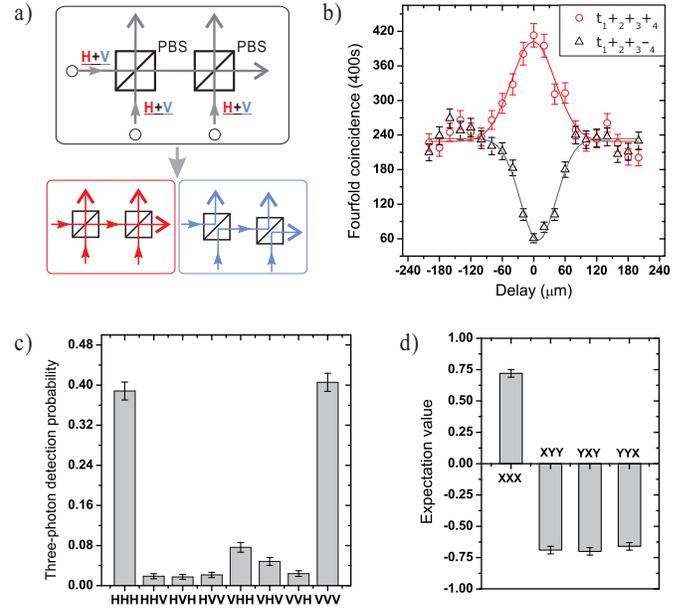}\\
  \caption{Principle of three-photon
  interference and data of genuine triplet entanglement.
  (a) Hong-Ou-Mandel type \cite{hom} interference of three photons. Three $|+\rangle$ polarized photons
  are directed to two PBSs from three spatial modes. As the PBSs transmit $H$ and reflect $V$ polarizations,
  a coincidence detection of three outputs can only originates from either the case that
  all photons are transmitted or all reflected. 
  (b) Three-photon interference
  as a function of the temporal delay. Outside the coherent regime the terms $|H\rangle_2|H\rangle_3|H\rangle_4$ and
  $|V\rangle_2|V\rangle_3|V\rangle_4$ are distinguishable. At the zero delay
  as an optimal superposition of $|H\rangle_2|H\rangle_3|H\rangle_4$ and
  $|V\rangle_2|V\rangle_3|V\rangle_4$ is achieved, thus the $|+\rangle_2|+\rangle_3|+\rangle_4$
  events show a maximal enhancement while the $|+\rangle_2|+\rangle_3|-\rangle_4$ events
  show a dip. The raw visibility is $0.75\pm0.03$, which
after subtraction of the contribution of the double pair emission,
can be improved to $0.88\pm0.05$, indicating good spatial and
temporal overlap have been achieved. This is, to our knowledge, the
first observation of Hong-Ou-Mandel type interference
  which involves three photons from three different paths. (c) The multiphoton detection probabilities in
  the $H/V$
  measurement basis. (d) Measured expectation value of the observables $XXX$, $XYY$, $YXY$ and $YYX$.
  The error bars denote
  one standard deviation, deduced from propagated Poissonian
  counting statistics of the raw detection events.
     }\label{}
\end{figure}

Now we move to the next step---execution of the semi-classical QFT,
as illustrated in Fig.~1c. We measure all possible correlations in
the state of the qubit $1$ and $2$, which is the simplest way to
simulate feedforward as used in Ref.~\cite{walther}. In the case
photon $1$ is measured in the state $|1\rangle$, an additional
quarter wave plate is inserted in the arm $2$ to implement the
$\pi/2$ rotation. Each measurement is flagged by a fourfold event
where all four detectors
fire simultaneously. 
The experimental results are shown in Fig.~4. With a probability of
$\sim50\%$ the output is in $|00\rangle$ corresponding to a failure.
Another $\sim50\%$ probability yields $|10\rangle$, which determines
the period $r=2^2/2=2,$ thus $\mathrm{g.c.d.}(11^{2/2}\pm1,15)=3,5$,
giving a successful factorization. To further quantitively evaluate
the performance, we use the squared statistical overlap \cite{fuchs}
of experimental data with the ideal values, which is defined as
$\gamma=(\Sigma^3_{y=0}m^{1/2}_ye^{1/2}_y)^2$, where $m_y$ and $e_y$
are the measured and expected output-state probabilities of the
state $|y\rangle$, respectively. From the data in Fig.~4 we find
$\gamma=0.99\pm0.02$, indicating a near perfect experimental
accuracy.
\begin{figure}[tb]
\centering
  \includegraphics[width=0.38\textwidth]{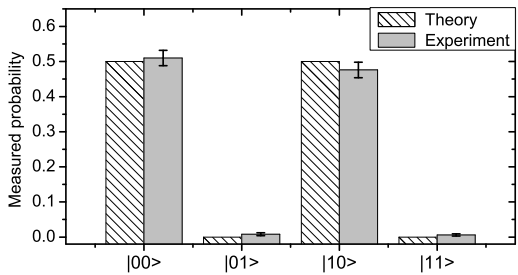}\\
  \caption{Experimental result of the output state probabilities after
  application of the QFT on the first register. Each measurement takes $120\mathrm{s}$
  and yields a maximum $270$ fourfold coincidence counts for the $|00\rangle$ projection.
     }\label{}
\end{figure}


It is noticeable that the performance of the algorithm is
considerably better than the quality of the entanglement created
between the registers. The imperfections of our experiment mainly
arise from high-order photon emissions and partial
distinguishability of independent photons \cite{gisin}, which cause
undesired mixtures in the GHZ state and degrade its fidelity.
However, for the execution of the algorithm, such mixtures happen to
have the same effect as the desired mixtures which could have been
resulted from the ideal circuit anyway (after the qubits $2$,~$3$,
and $4$ are entangled in the GHZ state, the qubit $2$ is in a
complete mixture tracing out of the qubit $3$ and $4$).

Some further remarks are warranted here. In this experiment we have
used the simplified optical two-qubit PBS gates which are
probabilistic and upon postselection \cite{dik}, thus scalability is
not directly implied in the present work. However, they can in
principle be improved to be deterministic using the scheme by Knill
\textit{et al.} \cite{KLM} given more photon source and complicated
linear optics network. Furthermore, an alternative approach, known
as the cluster-state model \cite{cluster,hein} provides a more
efficient realization of scalable photonic quantum computation
\cite{nielsen,walther,clusterexps}. In this model, universal quantum
computation is achieved by single-qubit measurements on a prepared
highly entangled cluster state.

Above we described a number of steps by which our MEF circuit was
brought to a simpler form. While these steps may seem at first
sight \textit{ad hoc}, in fact they are an example of a general
simplification method in the cluster state model. It has been
shown that  measurements of Pauli operator observables on cluster
states transform the state via a set of simple (and
computationally efficient) rules \cite{hein} to a  cluster state
with a different graphical description. Classical pre-processing
can thus produce a cluster state implementation with a smaller
number qubits, which, as demonstrated here, can simplify the
experimental realization. Thus, our approach can be seen as a
hybrid of cluster-state based and circuit-based models of quantum
computation,  adopting the most suitable model for the
implementation of MEF and QFT circuits respectively.

In summary, we have completed a proof-of-principle demonstration of
a complied version of Shor's algorithm using photonic qubits.
Genuine multiparticle entanglement is observed during computation,
which proves its unambiguous quantum nature. Our experiment presents
an important step toward full realization of Shor's algorithm and
scalable linear optics quantum computation. To factorize larger
numbers in the future, significant challenges ahead may include
coherent manipulations of more qubits, constructions of complex
multiqubit gates and quantum error correction \cite{chuang}.

We are grateful to L.M.K. Vandersypen, A. White and B. Zhao for
helpful discussions. This work was supported by the NNSF of China,
CAS, National Fundamental Research Program (under Grant No.
2006CB921900), the Alexander von Humboldt Foundation and Marie Curie
Excellence Grant of the EU. This work was also supported by Merton
College, Oxford and the EPSRC's QIPIRC programme.

\textit{note added}: --After submission of our manuscript, we became
aware of a related work \cite{white}.\\
Email $^\clubsuit$Chao-Yang Lu: cylu(at)mail.ustc.edu.cn;\\ Email $^\spadesuit$Dan Browne: d.browne(at)ucl.ac.uk;\\
Email $^\P$Jian-Wei Pan: pan(at)ustc.edu.cn.


\begin{thebibliography}{00}

\bibitem{shor1}
P. Shor, in Proc. 35th Annu. Symp. on the Foundations of Computer
Science (ed. Goldwasser, S.) 124-134 (IEEE Computer Society Press,
Los Alamitos, California, 1994).

\bibitem{shor2}
P. Shor, SIAM J. Comput. \textbf{26}, 1484 (1997).

\bibitem{RMP96}
A. Ekert and R. Jozsa, Rev. Mod. Phys. \textbf{68}, 733 (1996); E.
Gerjuoy, Am. J. Phys. \textbf{73}, 521 (2005).

\bibitem{rsa}
The RSA algorithm is named after R. Rivest, A. Shamir, and L.
Adleman who invented it in 1977. It is one of the most commonly used
public key systems nowadays (see http://en.wikipedia.org/wiki/RSA).

\bibitem{LMKV}
L.M.K. Vandersypen \textit{et al.}, Nature \textbf{414}, 883
(2001);\,\, L.M.K. Vandersypen and I.L. Chuang, Rev. Mod. Phys.
\textbf{76}, 1037 (2004).

\bibitem{nature}
S.L. Braunstein \textit{et al.}, Phys. Rev. Lett. \textbf{83}, 1054
(1999); N. Linden and S. Popescu, Phys. Rev. Lett. \textbf{87},
047901 (2001); R. Jozsa \textit{et al.}, Proc. R. Soc. A
\textbf{459}, 2011 (2003); G. Vidal, Phys. Rev. Lett. \textbf{91},
147902 (2003).

\bibitem{KLM}
E. Knill \textit{et al.}, Nature \textbf{409}, 46 (2001).

\bibitem{nielsen}
M.A. Nielsen,  Phys. Rev. Lett. {\bf 93}, 040503 (2004); D.E. Browne
\textit{et al.}, Phys. Rev. Lett. {\bf 95}, 010501 (2005).

\bibitem{rmp}
P. Kok \textit{et al.}, Rev. Mod. Phys. \textbf{79}, 135 (2007).

\bibitem{steinberg}
M. Mohseni, \textit{et al.}, Phys. Rev. Lett. \textbf{91}, 187903
(2003).

\bibitem{tame}
M.S. Tame, \textit{et al.}, Phys. Rev. Lett. \textbf{98}, 140501
(2007).

\bibitem{kwiatgrover}
P.G. Kwiat, \textit{et al.}, J. Mod. Opt \textbf{47}, 257 (2000).

\bibitem{walther}
P. Walther \textit{et al.}, Nature \textbf{434}, 169 (2005).

\bibitem{robert}
R. Prevedel \textit{et al.}, Nature \textbf{445}, 65 (2007).

\bibitem{preskill}
D. Beckman \textit{et al.}, Phys. Rev. A \textbf{54}, 1034 (1996).

\bibitem{Deutsch}
D. Deutsch, Proc. R. Soc. London Ser. A \textbf{400}, 96 (1985).

\bibitem{vedral}
V. Vedral, \textit{et al.}, Phys. Rev. A \textbf{54}, 147 (1996).

\bibitem{Lie-thesis}
L.M.K. Vandersypen, Ph.D thesis, Standford University (2001).

\bibitem{1cnot}
J.L. O'Brien \textit{et al.}, Nature \textbf{426}, 264 (2003).

\bibitem{dik}
This method has been used in quantum optics experiments where the
projection and the exploration of quantum states could be performed
simultaneously, see e.g., D. Bouwmeester \textit{et al.} Nature
\textbf{390}, 575 (1997).

\bibitem{griffiths}
R.B. Griffiths \textit{et al.}, Phys. Rev. Lett. \textbf{76}, 3228
(1996).

\bibitem{Kwiat}
P.G. Kwiat \textit{et al.}, Phys. Rev. Lett. \textbf{75}, 4337
(1995).

\bibitem{zukowski}
M. Zukowski \textit{et al.}, Ann. NY Acad. Sci. \textbf{755}, 91
(1995).

\bibitem{ghz}
D.M. Greenberger \textit{et al.}, Am. J. Phys. \textbf{58}, 1131
(1990).

\bibitem{hom}
C.K. Hong \textit{et al.}, Phys. Rev. Lett. \textbf{59}, 2044
(1987).

\bibitem{witness-exp}
C.A. Sackett \textit{et al.}, Nature \textbf{404}, 256 (2000); \,M.
Seevinck \textit{et al.}, Phys. Rev. A \textbf{65}, 012107 (2002);\,
M. Bourennane \textit{et al.}, Phys. Rev. Lett. \textbf{92}, 087902
(2004).


\bibitem{fuchs}
C.A. Fuchs, Ph.D. thesis, Univ. of New Mexico (1996).

\bibitem{gisin}
V. Scarani \textit{et al.}, Eur. Phys. J. D \textbf{32}, 129
(2005);\quad\quad\,\,\, M. Barbieri, Phys. Rev. A \textbf{76},
043825 (2007).

\bibitem{cluster}
R. Raussendorf \textit{et al.}, Phys. Rev. Lett. \textbf{86}, 5188
(2001).

\bibitem{hein}
M. Hein \textit{et al.}, Phys. Rev. A \textbf{69}, 062311 (2004).

\bibitem{clusterexps}
N. Kiesel \textit{et al.}, Phys. Rev. Lett. {\bf 95}, 210502 (2005).
\quad\,\, C.-Y. Lu \textit{et al.}, Nature Phys. {\bf 3}, 91 (2007).

\bibitem{chuang}
I.L. Chuang \textit{et al.}, Science \textbf{270}, 1633 (1995).

\bibitem{white}
B.P. Lanyon \textit{et al.}, Phys. Rev. Lett. \emph{to appear}
(2007).
\end{thebibliography}
\end{document}